\documentclass[doc]{apa}

\usepackage{graphicx} %
\usepackage{xspace} %
\usepackage{color} %
\usepackage{amsmath}
\usepackage{amsfonts}
\usepackage{amssymb}
\usepackage{paralist}
\usepackage[labelfont={sc}]{subfig}

\usepackage{setspace}
\usepackage{pifont}

\usepackage[us]{datetime}
\usepackage{fancyhdr}

\newcommand{\eg}{e.g.,\xspace} %
\newcommand{\ie}{i.e.,\xspace} %
\newcommand{\pd}{{\partial}} %
\newcommand{\Real}{\mathbb{R}} %

\newcommand{\eq}{Eq.\@\xspace} %
\newcommand{\eqs}{Eqs.\@\xspace} %

\title{Uncertainty of visual measurement\\ \vspace{.1in} and efficient allocation of sensory resources}
\author{\large Sergei Gepshtein$^{1}$ and Ivan Tyukin$^{2}$}
\affiliation{\vspace{.1in} $^1$The Salk Institute for Biological Studies, USA\\
\vspace{.1in} $^2$Leicester University, UK}

\ccoppy{Copyright: Gepshtein and Tyukin, 2014}

\abstract{\setstretch{1.05} \vspace{.1in}
We review the reasoning underlying  two approaches to combination of sensory uncertainties.  First approach is noncommittal, making no assumptions about properties of uncertainty or parameters of stimulation.  Then we explain the relationship between this approach and the one commonly used in modeling  ``higher level" aspects of sensory systems, such as in visual cue integration, where assumptions are made about properties of stimulation. The two approaches follow similar logic, except in one case maximal uncertainty is minimized, and in the other  minimal certainty is maximized. Then we demonstrate how optimal solutions are found to the problem of resource allocation under uncertainty.}

\begin{document}

\note{\vspace{.1in} \today}

\maketitle

{\setstretch{1.1} \tableofcontents}

\renewcommand{\theequation}{S\arabic{equation}}
\setcounter{equation}{0}

\setstretch{1.0}
\newpage
\section{1.~Combination of uncertainties}

\subsection{1.1.~Noncommittal approach}
\label{sec:bot_up_u}

\noindent Let the stimulus be an integrable function of one variable $I(x)$
that depends on two aspects of stimulation:
\begin{itemize}
  \item Stimulus location on $x$, where $x$ can be  space or time, the ``location" indicating, respectively \emph{where} or \emph{when} stimulation occurred.
  \item Stimulus content on $f$, where $f$ can be spatial or temporal frequency of stimulus modulation.
\end{itemize}
We consider a sensory system equipped with many measuring devices, each able to estimate both stimulus location and content  from $I(x)$. We assume that the error of estimation is a random variable with probability density $p(x,f)$.

It is sometimes assumed that sensory systems know $p(x,f)$: a case we review in the next section. But in general we do not know $p(x,f)$; we only know (or guess) some of its properties, such as its mean value and variance. In particular, let
\begin{equation}\label{sec:means}
\begin{split}
p_x(x) & =\int p(x,f)df, \\
p_f(f)  & = \int p(x,f)dx
\end{split}
\end{equation}
be the (marginal) means of $p(x,f)$ on dimensions $x$ and $f$. Sensory systems can optimize their performance with this minimal knowledge, as follows.

To reduce the chances of making gross errors, we use the following strategy.  We find the condition of \emph{minimal} uncertainty against the profile of \emph{maximal}
uncertainty, \ie using a minimax approach \cite{vonNeumann1928,LuceRaiffa57}. We do so in two steps. First we find such $p_x(x)$ and $p_f(f)$ for which measurement uncertainty is maximal. Then we find the condition at which the function of maximal uncertainty has the smallest value: the minimax point.

We evaluate maximal uncertainty using the well-established
definition of entropy \cite{Shannon1948}:
\begin{equation*}\label{eq:entropy:1}
H(X,F)=- \int p(x,f)\log p(x,f) dx \, df.
\end{equation*}
Recall that Shannon's entropy is sub-additive:
\begin{equation}\label{eq:entropy:2}
H(X,F)\leq H(X)+H(F)=H^{\ast}(X,Y),
\end{equation}
where
\begin{equation*}\label{eq:entropy:3}
\begin{split}
H(X) =  & - \int p_x(x)\log p_x(x) dx,  \\
H(F) =  & - \int p_f(f)\log p_f(f) df.
\end{split}
\end{equation*}
Therefore, we  can say that the uncertainty of measurement cannot exceed
\begin{equation}\label{eq:entropy:worst}
\begin{split}
H^{\ast}(X,F)  =  & - \int p_x(x)\log p_x(x)dx \\
                              & - \int p_f(f)\log p_f(f) df.
\end{split}
\end{equation}
\eq~\ref{eq:entropy:worst} is the ``envelope" of maximal measurement uncertainty: a ``worst-case" estimate.

By the Boltzmann theorem on maximum-entropy probability
distributions \cite{CoverThomas2006}, the maximal entropy of probability densities with
fixed means and variances is attained when the functions are
Gaussian. Then, maximal entropy is a sum of their
variances \cite{CoverThomas2006}. We obtain
\begin{equation*}\label{eq:entropy:4}
\begin{split}
p_x(x) & =\frac{1}{\sigma_x \sqrt{2\pi}} e^{-x^2/2\sigma_x^2},  \\
p_f(f)  & =\frac{1}{\sigma_f \sqrt{2\pi}} e^{-f^2/2\sigma_f^2},
\end{split}
\end{equation*}
where $\sigma_x$ and $\sigma_f$ are the standard deviations. And the
maximal entropy is simply:
\begin{equation} \label{eq:sum:variances}
H = \sigma_x^2 + \sigma_f^2.
\end{equation}
That is, when variances are unknown, maximal uncertainty of measurement is a sum of variances of measurement components.

This is the method used in the present derivations of \emph{joint uncertainty} and \emph{composite uncertainty} functions.\footnote{For simplicity,  we use intervals of measurement, rather than interval variances, as estimates of component uncertainties.}
The we find the optimal conditions by looking for minimal values of the uncertainty functions.

\subsection{1.2.~Top-down approach}
\label{sec:top_down_u}

\noindent Now we assume the system enjoys some knowledge of stimulation, so we can use \emph{likelihood} as a measure of uncertainty. Suppose we want to derive a combined estimate $z$
from two estimates $x$ and $f$ of some parameter of stimulation.
We assume that likelihood functions
$P(z|x,f)$, $P_x(z|x)$, and $P_f(z|f)$ are continuous,
differentiable, and known. Let us first assume that likelihoods are
separable:
\begin{equation}\label{eq:likelihood_trivial}
P(z|x,f)=P_x(z|x)P_f(z|f).
\end{equation}
Then, the most likely value of $z$ is
\[
z^\ast= \arg \max_{z} P(z|x,f)
= \arg \max_{z} [\log P_x(z|x) + \log  P_f(z|f)].
\]
We can use the logarithmic transformation because it is a strictly
monotone continuous function on $(0,\infty)$, and hence it does not
change maxima of continuous functions.

It is commonly assumed that $P_x(z|x)$ and $P_f(z|f)$ are Gaussian
functions, or that they are well approximated by Gaussian
functions. For example, \citeA{YuilleBuelthoff1996} assumed
that cubic and higher-order terms of the Taylor expansion of $\log
P_x(z|x)$ can be neglected, which is equivalent to the assumption of
Gaussianity. (We return to this assumption, and also the
assumption of separability in  a moment.)
Then
\begin{equation*}
\begin{split}
P_x(z|x)  & = c_x e^{-(z-z_x)^2/2\sigma_x^2}, \\
P_f(z|f)   & = c_f e^{-(z-z_f)^2/2\sigma_f^2}, \  c_x,c_f\in\Real_{>0}
\end{split}
\end{equation*}
and
\begin{equation*}
\begin{split}
&\log P_x(z|x) + \log  P_f(z|f)  = \\
& \log c_x + \log c_f -
\frac{1}{2\sigma_x^2} (z-z_x)^{2}- \frac{1}{2\sigma_f^2}
(z-z_f)^{2}.
\end{split}
\end{equation*}
The latter expression is maximized when its first derivative over
$z$ is zero. Hence
\begin{equation}\label{eq:p2}
\begin{split}
z^{\ast}
&=
\left(\frac{1}{\sigma_x^2}+\frac{1}{\sigma_f^2}\right)^{-1}\left(\frac{1}{\sigma_x^2}
z_x + \frac{1}{\sigma_f^2} z_f\right) \\
& = \frac{1}{\sigma_x^2 + \sigma_f^2} \, (\sigma_f^2  z_x + \sigma_x^2 z_f),
\end{split}
\end{equation}
which is the familiar weighted-average rule of cue combination
\cite{Cochran1937,MaloneyLandy1989,ClarkYuille1990,LandyEtAl95,YuilleBuelthoff1996}.
In general, when the number of measurements is greater than two, the combination rule of
\eq~\ref{eq:p2} becomes
\begin{equation}
z^{\ast}= \frac{1}{\sum_{i }\sigma_i^2} \sum_{i} z_i \prod_{j\neq i}
\sigma_j^2,
\end{equation}
where $z_i$ are such that individual likelihood functions attain
their maxima at $z_i$.

Why is the assumption common that likelihood functions have the simple
form of \eq~\ref{eq:likelihood_trivial}, \ie are separable and
Gaussian?
An answer follows from the argument we presented in the previous section.
Suppose that one seeks to estimate the likelihood
function when its shape is unknown.
We saw in the previous section that the
\emph{least certain} estimate is the
likelihood function for which the entropy is maximal. Hence, by
sub-additivity of entropy (\eq~\ref{eq:entropy:2}), the
least certain estimate of $P(z|x,f)$
is \[P(z|x,f)=P_x(z|x)P_f(z|f),\] as in
\eq~\ref{eq:likelihood_trivial}. Moreover, if the mean values and
variances of $P_x(z|x)$ and $P_f(z|f)$ are fixed, then the
likelihood functions must be Gaussian, by the same argument.
Indeed, separable Gaussian likelihood functions are the least certain estimates.

\section{2.~Resource allocation}

\noindent  We ask how sensory system ought to allocate their resources in face of uncertainties inherent in measurement and stimulation. We approach this problem in two steps.  First, we combine all uncertainties in \emph{uncertainty functions}: comprehensive descriptions of how quality of measurement varied across conditions of measurement. Second, we propose how limited resources are to be allocated given the uncertainty functions. Here we illustrate the second step in more detail, using the approach of constrained optimization.

A key requirement of allocation is  to optimize reliability (reduce uncertainty) of measurement by many sensors. Satisfying  this requirement alone makes the system place all sensors where conditions of measurement are least uncertain, leaving the system unprepared for sensing the stimuli that are useful but whose uncertainty is high. To prevent such gaps of allocation, we propose that minimal requirements should be twofold:
\vspace{0.1in}
\begin{description}
\item[Requirement A] Reliability: prefer low uncertainty.
\item[Requirement B] Comprehensiveness: measure all useful stimuli.
\end{description}

\vspace{.1in} \noindent We formalize these requirements as follows. Let:
\begin{itemize}
  \item $\Delta\in[a,b]\subset\Real$ be the size of measuring device (``receptive field"),
  \item $U(\Delta):\Real\rightarrow\Real$ be the uncertainty function associated with measuring devices
of different size, and
  \item $r(\Delta):\Real\rightarrow\Real_{\geq 0}$ be the amount of
resources allocated across $\Delta$ (\eq the number
of cells with receptive fields of size $\Delta$).
\end{itemize}

%\vspace{.1in} \noindent
%\textbf{Encouraging reliability.} \hspace{.025in}

\subsection{Encouraging reliability}

\noindent By requirement~A, the system is penalized for allocating resources where uncertainty is high. This is achieved, for example, when the cost for placing resources at
$\Delta$ is
\[k_1 U(\Delta)r(\Delta),\]
where $k_1$ is a positive constant.  The higher the uncertainty at $\Delta$,  or the
larger the amount of resources allocated to $\Delta$, the
higher the cost. Hence the total cost of allocation is:
\begin{equation}\label{eq:J1}
J_1=  \int_a^b k_1U(\Delta)r(x)d\Delta.
\end{equation}
Functional $J_1$ is minimal when all the detectors are allocated to (\ie have the size of) $\Delta$ at the lowest value of $U(\Delta)$.

%\vspace{.1in} \noindent
%\textbf{Encouraging comprehensiveness.}  \hspace{.025in}

\subsection{Encouraging comprehensiveness}

\noindent By requirement~B, the system
is penalized for failing to measure particular stimuli.
This is achieved, for example, when the allocation cost is
\[\frac{k_2}{r(\Delta)},\] where $k_2$ is a positive constant. The total penalty of
this type is:
\begin{equation}\label{eq:J2}
J_2=\int_a^b \frac{k_2}{r(x)}dx.
\end{equation}
Functional $J_2$ is large (infinite) when all resources are
allocated to a small vicinity (one point). $J_2$  is small when
$r(\Delta)$ are large for all $\Delta$.

%\vspace{.1in} \noindent
%\textbf{Prescription of allocation.} \hspace{.025in}

\subsection{Prescription of allocation}

\noindent The total penalty of requirements~A and B is
\begin{equation}\label{eq:J_total}
J=\int_{a}^b k_1 U(\Delta) r(\Delta) + \frac{k_2}{r(\Delta)}d\Delta.
\end{equation}
Using standard tools
of calculus of variations \cite<\eg>{Elsgolc1961}
we find such function $r(\Delta)$ that minimizes $J$. In particular, we consider a variation of $J$ with respect to changes of $r(\Delta)$:
\begin{equation*}
\begin{split}
\delta J=\int_{a}^b \frac{\pd }{\pd r(\Delta)}\left(k_1 U(\Delta) r(\Delta) + \frac{k_2}{r(\Delta)}\right)\delta r(\Delta) d\Delta& \\
             =\int_{a}^b \left(k_1 U(\Delta) - \frac{k_2}{r^2(\Delta)}\right)\delta r(\Delta) d\Delta&.
\end{split}
\end{equation*}
Because at optimal $r(\Delta)$ the value of $\delta J$ is zero for all $\delta r(\Delta)$, we deduce that conditions of
optimality are:
\begin{equation}\label{eq:allocation:1}
U(\Delta) - \frac{k}{r^2(\Delta)}=0,  \ k=\frac{k_2}{k_1}
\end{equation}
In other words
\begin{equation}\label{eq:r_or_delta}
r(\Delta)=\sqrt{\frac{k}{U(\Delta)}}.
\end{equation}
This $r(\Delta)$ is the prescription of  optimal allocation.

%\vspace{.1in} \noindent
%\textbf{Amount of resources.} \hspace{.025in}

\subsection{Amount of resources}

\noindent If the total amount or resources in the  system is known and is $C$:
\begin{equation}\label{eq:allocation:2}
\int_{a}^b r(\Delta) = C,
\end{equation}
then we may modify coefficients $k_1$ and $k_2$ in \eq~\ref{eq:J_total}, to make
\eq~\ref{eq:J_total} consistent with \eq~\ref{eq:allocation:2}. Or, we may use the method of Lagrange multipliers, looking for conditions where variation of the following functional vanishes:
\begin{equation}\label{eq:J_total_with_C}
\bar{J}=\int_{a}^b k_1 U(\Delta) r(\Delta) + \frac{k_2}{r(\Delta)}d\Delta + \lambda \left(\int_{a}^b r(\Delta) - C\right).
\end{equation}
We find Lagrange multiplier $\lambda$ at which \eq~\ref{eq:allocation:2} is satisfied. The solution (using a method similar to that used for solving \eq~\ref{eq:allocation:1}) is:
\begin{equation}\label{eq:allocation:3}
(k_1 U(\Delta) + \lambda) -\frac{k_2}{r^2(\Delta)}=0 \
\Rightarrow r(\Delta)=\sqrt{\frac{k_2}{k_1 U(\Delta)+\lambda}}
\end{equation}
provided that
\[
\int_{a}^b\sqrt{\frac{k_2}{k_1 U(\Delta)+\lambda}}d\Delta=C.
\]
\noindent The latter constraint is used to find  $\lambda$ in \eq~\ref{eq:allocation:3}.
In either case, the shape of the optimal allocation function $r(\Delta)$ is determined by $U(\Delta)$, such that
allocation function is maximal where $U(\Delta)$ is minimal.
The formulation in \eq~\ref{eq:J_total_with_C} has an advantage. It allows one to derive optimal prescriptions under changes in the amount of resources allocated to the task, such as in selective attention.

%\vspace{.1in} \noindent
%\textbf{Generalizations.} \hspace{.025in}

\subsection{Generalization to multiple dimensions}

\noindent In a multidimensional case, when $\Delta$ represents several variables (e.g., spatial and temporal extents of receptive fields, $S$  and $T$), and $U(\cdot)$  is a function of many variables, the prescription is
\[
r(s,t)=\sqrt{\frac{k}{U(s,t)}}.
\]

Using the method of Lagrange multiplies, one can show that a similar result is obtained when the costs of reliability and comprehensibleness (\eqs~\ref{eq:J1}--\ref{eq:J2}) have more general formulations:
\begin{equation*}
\begin{split}
J_1 &=\int_{a}^b k_1 U(\Delta) r^p(\Delta)d\Delta, \\
J_2 &=\int_{a}^b k_2 \frac{1}{r^{q}(\Delta)}d\Delta, \ p,q\geq 1,
\end{split}
\end{equation*}
The previously derived prescription holds: allocate maximal amount of resources to conditions of
minimal uncertainty.

\newpage
\setstretch{1.3} {\bibliography{supp}}

\end{document}